\documentclass [12pt,twoside]{article}   
 \usepackage[left,running]{lineno} 
\usepackage{setspace}

\countdef\arxiv=201

\ifnum\arxiv=0 \input cpreb.tex \fi

\ifnum\arxiv=1

\usepackage{epsfig,times,lscape}
\usepackage[usenames]{color}

    \ifnum\count102=1

    \fi

    \ifnum\count102=2
\fi

\newcommand{\nc}{\newcommand}

     \ifnum\count101=1
\nc{\qI}[1]{\section{{#1}}}
\nc{\qA}[1]{\subsection{{#1}}}
\nc{\qun}[1]{\subsubsection{{#1}}}
\nc{\qa}[1]{\paragraph{{#1}}}

\def\qpar{\vskip 2mm plus 0.2mm minus 0.2mm}
\def\qL{\hfill \break}
     \fi 

      \ifnum\count101=2
 \nc{\qI}[1]{\parindent=0mm \vskip 8mm 
{\centerline{\LARGE \color{red}#1}}\vskip 3mm}
\nc{\qA}[1]{\vskip 2.5mm \noindent 
{{\bf\large\color{blue}  #1}} \vskip 1mm \parindent=0mm}
 \nc{\qun}[1]{\vskip 1mm \noindent {\sl \color{blue} #1 }\quad }

\def\qL{\hfill \break}
\def\qpar{\vskip 2mm plus 0.2mm minus 0.2mm}

      \fi

\def\qth{\vrule height 12pt depth 0pt width 0pt}
\def\qtb{\vrule height 0pt depth 5pt width 0pt}

\nc{\qfoot}[1]{\footnote{{#1}}}

      \ifnum\count101=1
\def\qbu{\hfill \par \hskip 6mm $ \bullet $ \hskip 2mm}
\def\qee#1{\hfill \par \hskip 6mm (#1) \hskip 2 mm}
      \fi
      \ifnum\count101=2
\def\qbu{\hfill \par \hskip 4mm $ \bullet $ \hskip 2mm}
\def\qee#1{\hfill \par \hskip 4mm (#1) \hskip 2 mm}
      \fi

\def\qparr{ \vskip 1.0mm plus 0.2mm minus 0.2mm \hangindent=10mm
\hangafter=1}

     \ifnum\count101=1 
 \def\qdec#1{\parindent=0mm\par {\leftskip=2cm {#1} \par}}
     \fi
     \ifnum\count101=2
  \def\qdec#1{\parindent=0mm \par {\leftskip=1cm {#1} \par}}
  
  \def\qcitb#1{\noindent \hbox to 102mm{\hfill \small #1} \vskip 1mm}
      \fi

 \def\qpages#1{\count102=0{\loop\advance\count102 by 1
 \null \vfill\eject \ifnum\count102<#1 \repeat}}

\def\qn#1{\eqno \hbox{(#1)}}

\def\qth{\vrule height 12pt depth 0pt width 0pt}
\def\qtb{\vrule height 0pt depth 5pt width 0pt}

\def\qv{\vskip 0.1mm plus 0.05mm minus 0.05mm}

\def\qhw{\hskip 1.5mm}
\def\qleg#1#2#3{\noindent {\bf \small #1\qhw}{\small #2\qhw}{\it \small #3}\qv }
\newcommand{\promille}{%
  \relax\ifmmode\promillezeichen
        \else\leavevmode\(\mathsurround=0pt\promillezeichen\)\fi}
\newcommand{\promillezeichen}{%
  \kern-.05em%
  \raise.5ex\hbox{\the\scriptfont0 0}%
  \kern-.15em/\kern-.15em%
  \lower.25ex\hbox{\the\scriptfont0 00}}

\fi

\begin{document}
\thispagestyle{empty}

\color{yellow} 
\hrule height 10mm depth 10mm width 170mm 
\color{black}

 \vskip -10mm   

\centerline{\bf \Large \color{blue} Cliophysics: A scientific 
               analysis of recurrent historical events}
\vskip 10mm

\centerline{\normalsize
Yuji Aruka$ ^1 $,
Belal Baaquie$ ^2 $,
Xiaosong Chen$ ^3 $,
Zengru Di$ ^4 $,
Beomjun Kim$ ^5 $,
}
\qL
\centerline{\normalsize
Peter Richmond$ ^6 $,
Bertrand M. Roehner$ ^7 $,
Qing-hai Wang$ ^8 $,
Yang Yang$ ^9 $}

\vskip 5mm
\large

\vskip 5mm
\centerline{\it \small Version of 5 December 2021}
\vskip 3mm

{\small Key-words: Experimental physics, historical events, 
                comparative analysis} 

\vskip 20mm

{\normalsize
1: Institute of Economic Research, Chuo University, Tokyo, Japan.\qL
Email: aruka@tamacc.chuo-u.ac.jp \qpar

2: INCEIF (International Centre for Education in Islamic Finance), 
The Global University of Islamic Finance, Lorong Universiti A, 
59100, Kuala Lumpur, Malaysia. \qL
Email: belalbaaquie@gmail.com \qpar
3: School of Systems Science, Beijing Normal University, China.\qL
Email: chenxs@bnu.edu.cn \qpar
4: School of Systems Science, Beijing Normal University, China.\qL
Email: zdi@bnu.edu.cn \qpar
5: Physics Department, Sungkyunkkwan University, Seoul, South
Korea.\qL
Email: beomjun@skku.edu \qpar
6: School of Physics, Trinity College Dublin, Ireland.\qL
Email: peter\_richmond@ymail.com \qpar
7: Institute for Theoretical and High Energy Physics (LPTHE),
Pierre and Marie Curie campus, Sorbonne University,
National Center for Scientific  Research (CNRS),
Paris, France. \qL
Email: roehner@lpthe.jussieu.fr\qpar
8: Department of Physics, National University of Singapore, Singapore. \qL
Email: qhwang@nus.edu.sg \qpar
9: Astrophysics Department, Beijing Normal University, China.\qL
Email: 757103802@qq.com \qpar

}
\vfill\eject
%

\large

{\bf Abstract}\qL
Named after Clio, the Greek goddess of history,
cliophysics is a daughter (and in a sense an extension) 
of econophysics. Like econophysics it
relies on the methodology of experimental physics.
Its purpose is to conduct a
scientific analysis of historical events.
Such events can be of sociological, political or economic
nature. In this last case cliophysics
would coincide with econophysics.\qL
The main difference between cliophysics and econophysics is that
the description of historical events may be qualitative as well
as quantitative. For the handling of qualitative accounts
cliophysics has developed an approach based on the identification
of {\it patterns}. To detect a pattern the main challenge
is to break the ``noise barrier''.
The very existence of patterns is 
what makes cliophysics possible and ensures its success.
Briefly stated, once a pattern is detected, it allows predictions
to be made. As the capacity to make successful predictions is
the hallmark of any science, it becomes easy to decide whether or
not the claim made in the title of the paper is indeed fulfilled.
\qL
A number of examples of clusters of similar events will be given
which should
convince readers that historical events can be simplified
almost at will very much 
as in physics. One should not forget
that physical effects are also subject to the environment. 
For instance,
if tried at the equator, the experiment of the Foucault pendulum
will fail. \qL
In the last part of the paper, we describe  
cliophysical investigations conducted over the past
decades; they make us confident
that cliophysics can be a valuable tool for decision 
makers.

\vfill\eject

\qI{Introduction}

\qA{Objectives}

In this paper we will try to convince our readers that it is
possible to analyze recurrent historical events scientifically.
It will be seen that this is a natural extension
of physics and econophysics. Historical phenomena are
characterized by high levels of background noise (in a sense
defined more precisely later on) but nonetheless it is possible
to find regularities and patterns. This is our first objective.
Mathematical models may come later. Regularities and 
patterns already give predictions and conjectures  
that can be tested, which is the
distinctive trait of science.
This lecture will be really successful
if it leads those of our readers who have an interest in history as
well as in science  to try this approach by themselves. 
A kind of ``training-ground'' is proposed in appendix A.

\qA{Recurrent events in physics}

At first sight the objective of the paper may seem overly 
ambitious but one should observe that
the title does not promise  a scientific analysis
of history in a general way. It focuses on recurrent events that is
to say historical events which repeat themselves in similar
form. It is well known that in science reproductibility is 
a key requirement. It may be objected that historical
episodes do never repeat themselves exactly in same form.
But neither do physics experiments. As already mentioned in
the abstract, if one tries the experiment of the Foucault
pendulum in Singapore it will fail%
\qfoot{At the North Pole the oscillation plane of the 
pendulum will be seen to make a full 360 degree rotation
in 24 hours, in Europe it would take some 35 hours and
in Singapore it would take an infinite length of time, meaning
that the oscillation plane will not rotate at all.}%
.
In other words, it means that in this experiment 
the pendulum cannot be considered as a closed system.
It is subject to an exogenous factor which cannot be 
eliminated. Needless to say, subatomic phenomena
will be even more affected by exogenous factors. In a sense
head-on collisions of protons in an accelerator
occur in many different ways. In such cases, in order to
ensure reproductibility, large numbers of collisions 
must be recorded and their outcomes classified
in diverse groups of reactions.
\qpar

Table 1 shows that in its historical development physics
was firstly successful in the study of phenomena in which 
the background noise is almost nil. The refraction of light,
the orbits of planets are subject to very little exogenous
factors. For such phenomena measurement errors are the only
source of variability, and this is
a factor which can be gradually reduced
as measurements become more accurate. 

%
\begin{figure}[htb]
\centerline{\psfig{width=15cm,figure=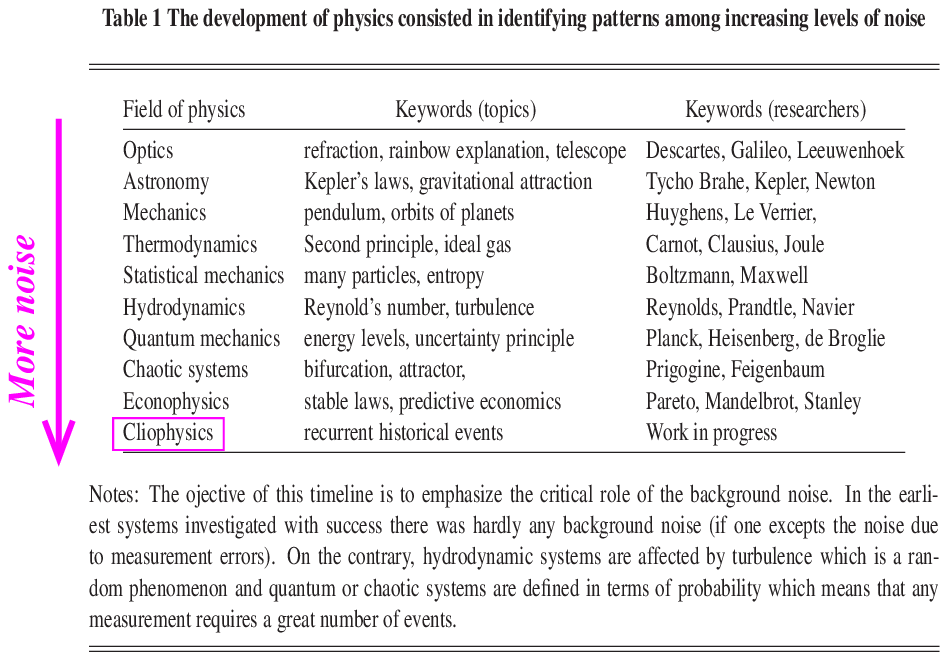}}
\qleg{Fig.1 Historical development of physics.}
{The first successes of modern physics (that is to say
not including progress made in ancient Greece) concerned
phenomena in which there was almost no background noise.
In contrast, hydrodynamical phenomena are affected 
by turbulence, subatomic phenomena are subject to
basic quantum uncertainty, chaotic phenomena are
very sensitive to initial conditions which is
referred to as the butterfly effect.}
{}
\end{figure}

Table 1 does not list all fields
of physics; for instance, crystallography, electromagnetism
and a number of others are omitted. It
focuses mostly on fields of physics which were developed
in the 20th and 21th centuries.
In hydrodynamics, quantum physics
and chaotic systems there is a high level of background noise,
that is to say exogenous, random factors.
This may come to the point, for instance in percolation
effects, that sometines it is  unclear whether or not an effect
is reproducible.

\qA{The spirit of physics}

All too often the approach of physics
is reduced to its mathematical formalism. This conception has
become prevalent particularly since 1900. However, the
spirit of physics can also teach us important lessons
for how to conduct experimental investigations. \qL
The first lesson, namely the reduction
of the background noise was already mentioned in the 
previous subsection.\qL
The second lesson is the simplicity requirement.
Instead of a ball rolling on a wooden plank,
Galileo could have studied the trajectory of cannon projectiles.
That would probably have raised the interest of his sponsors,
but the attempt would have ended in failure for the obvious
reason that this problem is much more difficult.\qL
Is it possible to consider simple cliophysics
``experiments''? We think so. 
\qpar

Let us consider the case of a pendulum. This is also a topic
in which Galileo was interested. More precisely, he studied
the period of a pendulum that is now called a simple pendulum.
It is a pendulum which moves in a steady vertical plane.
However, a spherical pendulum has at least half a dozen
different movements: the Puiseux effect, the Foucault effect, 
the added mass effect, and so on. If all these movements are
allowed to occur simulaneously, the movement becomes
complicated to the point of defying any modeling attempt.
How could physicists overcome this obstacle? \qL
The answer is simple. 
\qpar

For each of the different effects they designed a specific 
experiment which eliminated all other effects except the one
they wanted to consider. \qL
For instance, to observe  the
Foucault effect it is well known that one should use a
very long pendulum (50m or more) for otherwise the 
Puiseux effect will interfere with the Foucault effect
and will lead to fairly random measurements. \qL
In the same spirit to observe the added-mass effect
one should use a pendulum which moves in water rather than
in air for otherwise the added-mass effect will be 
too small compared to other effects.
\qpar

Can we use the same approach for social and historical
phenomena? We believe so. The challenge is to select
the events in which the effect that we wish to observe 
appears most clearly. How to do that will be explained
in the next subsections.
\qpar

Firstly, we explain how the background noise can be reduced.
Secondly, it will be explained on an example how to generate
events which can be defined by only 2 or 3 parameters.

\qA{Recurrent events in history}

We face the same difficulty as in physics
in the sense that each of the episodes will 
be subject to diverse exogenous factors in the form of 
background noise. Our clusters
of similar events may include episodes taking place in
different environments
(e.g. various  countries and centuries) which may translate
into fairly high background noise. \qL
How can we overcome such a  serious obstacle?
\qpar

Clearly, we should follow the example of physics.
Instead of trying to study the fairly complex episodes which 
interest historians (e.g. the American Revolution or
the French Revolution of 1789) we should focus on
very simple and basic cases in which there is little
background noise. We will list a number of cases of that
sort in a short moment. 
\qpar

It is almost certain that 
historians may find
such cases weird and uninteresting but this should
not stop us. After all, in the time of 
Tycho Brahe and Kepler very few people found any interest
in their accurate measurements. The field was dominated
by astrologers%
\qfoot{It should be remembered that Kepler himself 
was the official astrologer of the emperor of the 
Holly Roman Empire.}
for whom the accuracy of existing tables 
was sufficient.
\qpar

We should completely put aside any anthropocentric
attempts to understand human behavior in the way 
practised by historians. 
We should perform basic measurements of simple effects.
Little by little, through the magic of cross-fertilization,
our understanding will expand. At first sight it was not
obvious that the study of the refraction of light may help us to
understand the rainbow phenomenon, but nevertheless this is what 
happened.
\qpar

It is at this point that the 
role of physicists, and particularly of econophysicists,
will be crucial. Now that it is fairly clear that historians 
will neither approve nor be interested in cliophysics, the only
way this field may develop is through the cooperation of
physicists; in this term we include physical chemists, 
astrophysicists and also laymen with a taste for experimental
investigation.

\qI{Historical episodes suitable for cliophysical investigation}

\qA{Overall view}

As emphasized above we must focus on episodes with as little
background noise as possible. In this section we mention two
categories of cases.
\qbu Cases in which similarity in objectives and agents
ensures a relative uniformity and low variability.
The cases listed in Table 2
are of that kind.
\qL
However, when considered across nations and various time periods,
the events listed in Table 2 involve a sizeable number of 
parameters. In other words they are certainly less simple than
the free fall experiments performed by Galileo.
\qbu Just in order to show that one can find recurrent events
which depend upon even less parameters than those in Table 2,
we introduce the set of events listed in Table 3. We call them
extremely simple events; they are of the same degree of
simplicity as Galileo's free fall experiments. 
At first sight they may appear somewhat weird. The important point
is that they depend upon only 3 well defined yes--no parameters. 
It will be seen that despite their extreme simplicity, it
is nevertheless possible to draw some meaningful conclusions
from their comparison.

%
\begin{table}[htb]

\small

\centerline{\bf Table 2 Examples of low noise events}

\vskip 5mm
\hrule
\vskip 0.7mm
\hrule

$$ \matrix{
\qth
&\hbox{Event} \hfill & \hbox{Number of} & \hbox{Example} \hfill \cr
\qtb
&\hbox{} \hfill & \hbox{cases} & \hbox{} \hfill \cr
\noalign{\hrule}
\qth
1&\hbox{Peasants' revolt} \hfill & \sim 300 & 
\hbox{Wat Tyler revolt, England, 1381)} \hfill \cr
2&\hbox{Major mushroom strike} \hfill & \sim 20 & 
\hbox{France, May 1968} \hfill \cr
3&\hbox{General strike planned by union} \hfill & \sim 20  & 
\hbox{Switzerland, November 1918} \hfill \cr
4&\hbox{Rejection riot} \hfill & \sim 50 & 
\hbox{Tulsa, Oklahoma, 1921} \hfill \cr
5&\hbox{Protest riot} \hfill & \sim 100 & 
\hbox{Detroit, Michigan 1969} \hfill \cr
6&\hbox{Massacre of peaceful demonstrators} \hfill & \sim 50 & 
\hbox{St Petersburg, January 1905} \hfill \cr
7&\hbox{Mutiny on land} \hfill & \sim 50 & 
\hbox{Pennsylvania Line mutiny, 1783} \hfill \cr
8&\hbox{Mutiny on ship} \hfill & \sim 20 & 
\hbox{Hermione mutiny (Royal Navy ship,1797)} \hfill \cr
\qtb
9&\hbox{Prison riot} \hfill & \sim 100 & 
\hbox{Attica prison, New York State, Sep 1971} \hfill \cr
\noalign{\hrule}
} $$
\vskip 1.5mm
Notes: Note the distinction between ``mushroom strikes'' which
are fairly spontaneous movements by grassroot workers (closely
analyzed in Roehner and Syme 2002)
and general strikes organized by unions.
Note also the distinction between rejection and protest
riots. The former are aimed against a minority group
which is not accepted by the residents. A protest riot is directed
against the state. Some riots have a mixed status; for instance,
the Gordon riots in London (1780) were anti-Catholic but were
also directed against the government who had passed a law which
restored some of the rights of the Catholic minority.
Some of the events listed above (particularly 2,3,4) have been 
studied in Roehner and Syme (2002). The number of cases are 
understood to be at worldwide level but, needless to say, these
numbers give merely orders of magnitude.
\qL
\vskip 2mm
\hrule
\vskip 0.7mm
\hrule
\end{table}

\qA{Requirements for extremely simple events}

Events well suited for comparative analysis
must be as simple as possible (i.e. involve only few
parameters)
and in addition they should
satisfy the following requirements. 
\qbu The most important condition is that
these events must be well documented in all their 
aspects for otherwise any comparative analysis would 
be impossible. This means that they
should
attract sufficient  attention from contemporaries to
receive detailed accounts.
\qbu These events should not be limited to a specific country
but exist everywhere so as to allow comparison between
various countries and time periods.
\qpar

So far, the only events we have found that conform to these
requirements concern the execution of renowned dignitaries.
We beg our readers to pardon
the recourse to such macabre cases. Actually,
there is probably a logic in this, in the sense
that it is precisely because such events are gruesome
that they impressed contemporaries and were
duly recorded.
Naturally, that is all the more true when the executed
persons are
kings, queens or other well known dignitaries.
\qpar

Another virtue of this example is to press on the idea
that, as was stronly emphasized by sociologist Emile
Durkheim (1894), we should completely drop our anthropomorphic
vision. What matters here is not whether the topic
is macabre or not, but whether the cases are simple 
and well documented.

\qA{Protocol of execution of dignitaries}

Table 3 gives several examples of executions of renowned persons.
What can be the interest of such a list?\qL
The key-point is this. \qL
If considered individually (i.e. without any mention 
of parallel executions)
they do not raise any question. We can just accept each
description as a unique event. This makes it impossible to gain
any understanding. \qL
In contrast, their comparison brings about 
several questions.

Many accounts have been devoted to the life of
Mary Stuart, Queen of Scots and her trial during the reign
of Queen Elizabeth I. However, little attention has been given
to how she was executed. 
%
\begin{table}[htb]

\small

\centerline{Table 3 Pattern versus variability in the
execution of dignitaries}

\vskip 5mm
\hrule
\vskip 0.7mm
\hrule

$$ \matrix{
\qth
\hbox{Victim} \hfill & \hbox{Year} & \hbox{Status} \hfill& \hbox{Axe or} &
   \hbox{Head} & \hbox{Head} & \hbox{Hand} \cr
\qtb
\hbox{} \hfill & \hbox{} & \hbox{} & \hbox{Sword} &
   \hbox{displayed} & \hbox{exposed} & \hbox{cut} \cr
\noalign{\hrule}
\qth
\hbox{Marcus Cicero} \hfill & -43 & \hbox{Consul} \hfill& \hbox{S} &
   \hbox{no} & \hbox{yes} & \hbox{yes} \cr
\hbox{Anne Boleyn} \hfill & 1536  & \hbox{Queen} \hfill& \hbox{S} &
   \hbox{?} & \hbox{no} & \hbox{no} \cr
\hbox{Catherine Howard} \hfill & 1542 & \hbox{King's wife} \hfill& \hbox{A} &
   \hbox{?} & \hbox{no} & \hbox{no} \cr
\hbox{Jane Parker} \hfill & 1542  & \hbox{?} \hfill & \hbox{A-S} &
   \hbox{?} & \hbox{no} & \hbox{no} \cr
\hbox{Mary Stuart} \hfill & 1587 & \hbox{Queen} \hfill& \hbox{A} &
   \hbox{yes} & \hbox{no} & \hbox{no} \cr
\hbox{Charles I} \hfill & 1649 & \hbox{King} \hfill & \hbox{A} &
   \hbox{yes} & \hbox{no} & \hbox{no} \cr
\hbox{Oliver Cromwell} \hfill &  1661& \hbox{} & \hbox{S} &
   \hbox{yes} & \hbox{yes} & \hbox{yes} \cr
\qtb
\hbox{Johann Struensee} \hfill & 1772 & \hbox{Count} \hfill& \hbox{A} &
   \hbox{yes} & \hbox{yes} & \hbox{yes} \cr
\noalign{\hrule}
} $$
\vskip 1.5mm
Notes: From Cicero to Struensee the protocol of the execution 
was almost identical and served the same objective, namely
to destroy a once powerful political leader. Whereas in the case
of Oliver Cromwell the objective was also the same, the method 
was fairly different because in 1661 Cromwell was already 
dead which means that it was a posthumous execution. His body
had to be exhumed from Westminster Abbey. Subsequently to his
beheading his body was cut into pieces and his
head was displayed on a pole on the roof of Westminster Hall
until 1685. In the case of Jane Parker (who was the wife of the 
brother of Anne Boleyn) the two sources we have found (one in French,
the other in English) give conflicting information, hence the S--A 
mark. Head displayed means that the head was shown by
the executioner to the public. Head exposed means that
instead of being buried the head was displayed in a public
place. Thus, Cicero's head and hand were displayed in the Forum.
Note that Cicero was not tried but simply blacklisted by
his political opponents. Blacklisted persons could be
killed by anybody.
\qL
Sources: Fraser (1969) and biographical articles of Wikipedia.
\vskip 2mm
\hrule
\vskip 0.7mm
\hrule
\end{table}

%
\begin{figure}[htb]
\centerline{\psfig{width=10cm,figure=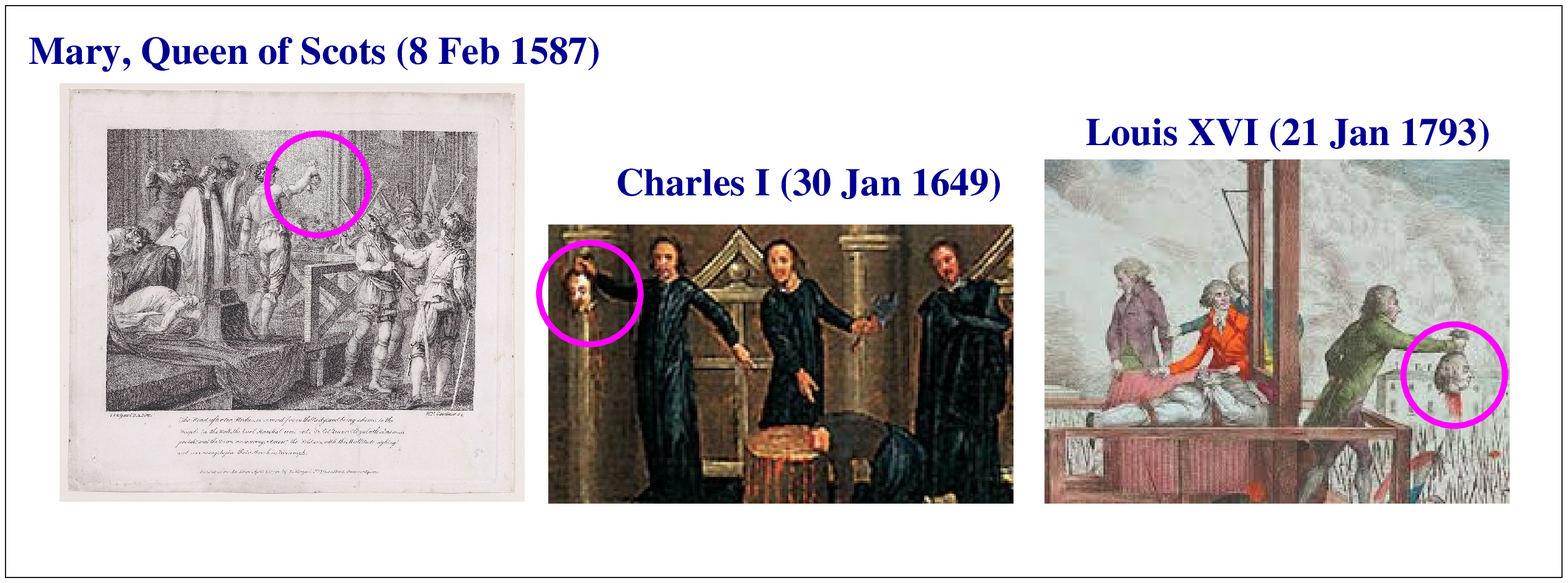}}
\qleg{Fig.2: A key-feature in the execution of queens,
kings and other high dignitaries was to show their head to
the population.}
{Historical records attest that the executioner was specifically
instructed to show the head while spelling
out the reason of the execution (e.g. ``this is the head of a
traitor''). Such events are particularly suited for comparative
analysis because their close similarity allows us to focus on
slight differences.}
{}
\end{figure}

In the case of Anne 
Boylen one can mention the following observations.
\qbu Anne Boleyn was the second wife of King Henry VIII. A recently 
discovered historical record (Alberge 2020) shows
that Henry VIII planned every detail of her beheading. 
\qbu  If one excludes
the dubious case of Jane Parker, Anne Boleyn was the only person in
our list who was beheaded with a sword. Some historians argued
that among all persons executed during the reign of Henry VIII
her case was unique. Clearly this is a point which is difficult
to prove because we do not have detailed information about
the numerous executions%
\qfoot{After the ``Pilgrimage of Grace'' which was a Catholic
 uprising, some 200 noblemen were executed.}
which occurred. At least it can be said that in a sense her execution was
set up with ``special regard''.
\qpar

More could be said about Table 1 but we will limit ourselves
to only one additional comment. It concerns the case 
of Count Struensee in Denmark. As Prime Minister he was an
adept of the Entlightening and he was also the lover of the queen.
Before being beheaded he had his right hand
cut off. Interpreting this as special cruelty would certainly be a
misconception for the case of Marcus Cicero shows that this
custom also existed in the Roman Republic. It is well known that
in the 18th century ancient Rome was still a model. This was
particularly true for Roman law and classical 
architecture.

\qA{Fate of vanquished enemies and traitors}

Beheading vanquished enemies 
or traitors has been practised worldwide
from ancient times to the 21st century. Many specific historical
cases are cited in Roehner (2007, p.56-57). The judicial executions
of dignitaries considered above should be seen as a formalized
subset of a much broader tradition. Ignorance of this
tradition will lead to the misunderstanding of many historical
occurrences. Here are three examples.
\qbu On the South Bank of  the London Bridge there is a tower
called Bridge Gate which has a flat top. On this place,
from 1300 to 1678,
there were dozens of heads of traitors (including those of Thomas
More and bishop John Fisher)
spiked on poles. After 1678 they
were moved to Temple Bar. For better preservation the heads
were dipped in tar.
\qbu After his victory in Sudan against the Mahdists 
at the battle of Omdurma (2 September 1898) English Major-General
Kitchtener (he would become Lord Kitchener only in 1914) had
the body of the Mahdi exhumed and his head sent to Cairo.
\qbu After the Bastille was taken by the people in the early days of the 
French Revolution%
\qfoot{The victory was made possible because in the confrontation 
the troops sided with the insurgents.}
the head
of governor de Launay was paraded at the top of a pole. Although
a familiar image of the Revolution it is not in the least specific to
this event. In fact this was a well established usage.
Naturally, historians whose interest is narrowly 
focused on the Revolution may not know that, and  
therefore cannot understand the real significance.
\qpar

By capturing attention this image may distract historians
from aspects  which were more specific to the Revolution.
One of these aspects is the important
fact that the insurgants had weapons.
As a matter of fact, the main reason for attacking the Bastille
was to take the powder and guns that were stored in the fort.
This is well attested by the fact that de Launay was asked to
deliver the stored weapons and it is only on his refusal
that hostilities started.
\qpar

It is only through comparison with other cases that 
one comes to realize
that it was not common for insurgents to be in possession of
guns and even cannons. A comparison with the Gordon riots of
June 1780 in London is quite revealing. Rud\'e (1955, 1964)
describes
the rioters as being armed with ``bludgeons, crowbars and chisels''.
Nevertheless and quite surprisingly, they were able to take
several of the main prisons, e.g. Newgate, Fleet prison, King's Bench
prison, New Goal prison. Why did the rioters not attack
armuries, naval stores and other places where they would find
arms? \qL
The Gordon riot was not an isolated case. The ``John Wilkes riot''
of 10 May 1768 saw the gathering of 15,000 demonstrators, all unarmed.
\qpar

In short, neglect of comparative analysis of the kind
promoted by cliophysics will result for historians 
in serious pitfalls: e.g.
failure to rightly interpret head spiking or
blindness to the fact that the Parisian insurgents were
armed (which was mainly due to the fact that the National
Guard was on their side).
\qpar

In the same line of thought, let us briefly mention another
case in which lack of a comparative perspective led
historians, and particularly French historians, to 
misjudgements. It concerns religious tolerance in France.
As one knows, after several decades of civil wars between 
Catholics and Protestants, in 1588 the Edict of Nantes opened
a path of secularism and tolerance. Naturally French Republican 
historians warmly approved King Henry IV for signing this edict.
\qpar

However, being focused on France, they do not realize
that this edict was an anomaly in Europe. 
All other European powers
(Britain, Germany, Poland, Spain, sweden) had introduced laws
favoring religious uniformity. The mildest of such laws closed
access to state employment to members of the religious minority.
There were many other laws which were more severe.
\qL
Moreover, in 1618, i.e. three decades after the introduction
of the Edict of Nantes,  started the ``Thirty Year War''
which opposed Catholics and Protestants in Germany and was
one of the most destructive European conflicts, at least before
the world wars.
\qpar

Then, in 1685 the Edict of Fontanebleau revoked the Edict of Nantes.
Needless to say, 
the revocation raised strong disapproval among French historians.
Yet, at this time, intolerance was still the most common
attitude in all nations; e.g. the Quakers were not
allowed in Puritan Massachusetts under the threat of death.
\qpar

What we wish to emphasize is that disapproval is an anthropomorphic
attitude which leads historians to misjudgements. The best
antidote is the comparative perspective promoted by cliophysics.

\qI{Testing cliophysical models} 

The hallmark of scientific analysis 
consists in the successful testing of laws. These laws,
we will argue here, can be quantitative as is usually the
case in physics but can also be qualitative as will often be 
the case in cliophysics. 
Qualitative laws state that, under a number 
of specific conditions, one shall observe a given effect.
In quantitative laws one will in addition be able to
predict the magnitude of this effect. 
\qpar

In this section we will explain how comparative analysis
naturally leads to qualitative predictions and how such predictions
can be tested. Firstly, in order to become more familiar
with this method we show how it can be used in physics.

\qA{How to use the comparative approach in physics?}

Suppose you are living in
the time of Galileo (1564--1642) and
Descartes (1596--1650) and you ask yourself
what are the forces involved in the phenomenon of free fall.
A first idea is to try Galileo's method of dropping 
balls of different sizes and densities from the tower of Pisa.
However, with the poor means of observation available in that
time, this method did not work very well. 
\qpar

Another approach is to try clever comparisons of extreme cases.
The expression ``extreme cases'' means that we do no longer compare
the fall of balls made of iron or wood but instead wish to try
balls whose densities are radically different, 
for instance by observing the
fall of (a) lead balls (b) soap bubbles (c) feathers. 
\qpar

As one knows, there are 3 forces at work. (i) the weight $ mg $
(ii) the buoyant force (iii) the friction of air. 
In (a) one sees almost only (i). In (b) the buoyant force (ii)
is magnified and by trying bubbles of
different diameters, one can identify its characteristics. 
In (c) it is the friction of air which is amplified. 
In his publications Galileo does not
report that he used such a comparative method. 
It would  have 
given him a better understanding of the different effects at work.

\qA{How to use comparative analysis in cliophysics? The case
of the Pacific}

As an illustration of how this method may work we consider
the ``zone of influence'' effect. 
\qpar

A country $ A $ may have a zone of influence $ Ab $ in a country
$ B $ when it has rights in $ Ab $ that it does not have in the rest
of $ B $. For instance, France had rights in the French concession
of Shanghai that it did not have in the rest of China.\qL
In an extension of this meaning, a zone of influence may consist in
a whole country. For instance, Egypt was a British zone of influence
until the end of World War II because it gave access to India,
the jewel of the British Empire. 
\qpar

The degree of influence ($ di $) that country $ A $ exercises in 
region $ Ab $ is variable on a scale from $ d-i=0\% $ when there is
no influence at 
all to $ d-i=100\% $ in case of a colony. In the case of Britain 
one may say that it had a 60\% influence in Egypt and
a 100\% influence in India, at least in  the parts of
India which were under direct British rule. 
\qpar

Clearly, it is a challenging task to determine the $ d-i $
of a given area. On 2 June 1954, in a ``Security Conference'',
President Eisenhower declared: ``We have got to keep the
Pacific as an American lake''. What is the implication in
terms of degree of influence?
\qpar

Ever since the United States took possession 
of the West Coast (the conquest of California was in 
1847 during the US-Mexican War) it has had an
hegemonic position in the Pacific Ocean. 
As a matter of fact,
from 1850 to around 2010 it was by far the most important
power of the Pacific rim. Therefore it is
quite understandable that in 1954 President
Eisenhover could called it an American lake. 
\qpar

In a book by Di et al. (2017) and in a 
paper by Baaquie et al. (2019) the authors try
to determine what is the US $ d-i $ over the Pacific rim
and what may be the implication for the future of the relations
between China and the US.\qL
US responses to successive 
encroachments upon its hegemony from 1880 to now were 
systematically studied. It turns out that during this
time interval the US has been unwilling to consider
a negotiated partnerships preferring to give a free
hand to its military%
\qfoot{Yet, one can remember that in April 1951
President Truman
reined in General McArthur's in his plans against China.
In this plan ``between 30 and 50 atomic bombs'' would
have been  dropped on China, see the interviews
of MacArthur taken in 1954 but which were 
published in the ``New York Times'' ten years later on 
9 April 1964 (p.16).}%
.

Now, in December 2021 four years have past since
the first publication and 2 years since the 
publication of the paper by Baaquie et al. Although
the future is still cloudy these past years confirmed
the fairly pessimistic predictions at least in the sense
that no progress has been made in the direction
of a compromise. 
\qpar

What kind of compromise can one think of? Here too 
the analysis of clusters of former similar episodes
comes to our help. It shows that zones of influence
belonging to different countries may co-exist. That is
what happened in China in the time of foreign concessions:
Japan and Russia in the north, Britain and France in the 
South and Germany in the east.
This arrangement created little friction between
foreign powers, at least until Japan started its 
invasion in 1933. The occupation of the whole western coast
by Japan completely destroyed the previous state of equilibrium.\qL
Another example of an agreement about respective zones of
influence occurred in Africa between Britain and France.
The co-existence of the two zones of influence did not
provoke much friction.
\qpar

The United States has already a near 100\% zone of influence
in South Korea, Japan and Australia and possibly a 60\% zone
of influence in India. Could that not be the basis of a fair
arrangement with China that would leave enough breathing space
to each side. Naturally, our prediction that things will
go from bad to worse is probabilistic; its interval of confidence
depends upon whatever new factors may appear which did not
exist in the former cases.

\qI{Conclusion and perspectives}

Cliophysics has had a difficult start because its early publications,
e.g. the book of 2002, 
were aimed at historians. With the benefit of hindsight it has
become clear that most historians do not accept the approach
of cliophysics. At the other end of the spectrum, because (so far)
it was expressed in qualitative rather than mathematical terms,
cliophysics did not attract the interest of econophysicists
or astrophysicists. Our advice is to firmy rely on the
spirit of physics. For several centuries it has guided
physicists in their experimental explorations and we believe it
will also be an excellent guide for cliophysical explorations.
\qpar

As nothing can replace personal experience, we describe in
Appendix A a test-study that can be tried by those of our readers
who are in sympathy with the approach and goals of cliophysics.
We are aware of the fact that the episodes described in this
appendix may involve too many parameters. Therefore, the first
challenge is to define simple sub-episodes which may be compared
in an effective way and lead to well-defined patterns. 
Simplicity is the key for success.

\appendix

\qI{Appendix A. The fate of Loyalists as a training-ground}

Those Americans who
sided with the British during the War of Independence
had their property confiscated
and eventually they emigrated to other parts of the
British Colonial Empire. The fate of such ``Loyalists''
has attracted the attention of historians.
\qpar

Needless to say, all great changes, whether political or
religious, produce Loyalists that is to say people who
wish to stick to former conditions.
When Sweden became a Protestant country in the 16th
century, the people who remained Catholic were ``Loyalists''.
As they did not enjoy the same rights than Protestant citizens,
instead of remaining in the country some of them
fled to Catholic countries%
\qfoot{Incidentally, in 2021 Protestantism remains the official State
Church in Denmark, Finland, Iceland and Norway but in Sweden
there is no official State Church currently.}%
.
The same observation can be made
in Britain where Puritans, Quakers, Catholics and members of
non-conformist denominations emigrated to America:
Puritans to Massachusetts, Quakers to Pennsylvania or New Jersey,
Catholics to Maryland.
\qpar

Apart from religious revolutions one can also
consider  political upheavals, e.g. the English Revolution,
the French Revolutions of 1789,
the Russian Revolution of 1917, the Chinese Revolution
of 1949, the end of the Algerian war of independence in 1962,
the recent change of regime in Afghanistan (2021),
all these cases produced Loyalists. Were they indicted by the
new regime? Did they have their property confiscated? 
How many were killed? How 
many emigrated? How many returned a few years later?
For all these questions the main difficulty is to find
reliable sources. However, thanks to the Internet, this task
has now become possible while it would have been
impossible prior to the Internet era.

\vskip 10mm

{\bf References}

\qparr
Alberge (D.) 2020: Chilling find shows how Henry VIII planned
every detail of Boleyn beheading. 
The Guardian 25 October 2020.

\qparr
Baaqui (B.E.), Wang (Q.-H.) 2018:
Chinese dynasties and modern China. Unification and fragmentation.
China and the World, Ancient and Modern Silk Road 1,1,1-43.

\qparr
Baaqui (B.E.), Richmond (P.), Roehner (B.M.), Wang (Q.-H.) 2019:
The future of US-China relations: a scientific investigation.
China and the World. Ancient and Modern Silk Road 2,1,1-53.

\qparr
Bloch (M.) 1924: Les rois thaumaturges. 
\'Etude sur le caract\`ere surnaturel attribu\'e \`a 
la puissance royale particuli\`erement en France et en 
Angleterre. Istra, Paris.\qL
Translated into English under the title: The royal touch.
Sacred monarchy and scrofula in England and France.
Routledge 1974.

\qparr
Di (Z.), Li (R.), Roehner (B.M.) 2017: 
USA-China: cooperation or confrontation. 
A case study in analytical history (in Chinese). \qL
[This book can be read at the following address:\qL
http://www.lpthe.jussieu.fr/$ ~ $roehner/prch.php]

\qparr
Durkheim (E.) 1894:  Les r\`egles de la m\'ethode sociologique.   
Flammarion,  Paris.\qL
[The book has been translated into 
English under the title:  \qL
``The rules of sociological method''.  \qL
Both the French and the English version are freely available 
on Internet.]

\qparr
Fraser (A.) 1969: Mary, Queen of Scots. 
Weidenfeld and Nicholson, London.

\qparr
Roehner (B.M.) 1997a: Jesuits and the state. A comparative 
study of their expulsions (1520-1990).
Religion 27,165-182.

\qparr
Roehner (B.M.) 1997b: Spatial determinants of separatism.
Swiss Journal of Sociology 23,1,25-59.

\qparr
Roehner (B.M.), Rahilly (L.) 2002: Separatism and integration.
A study in analytical history. 
Rowman and Littlefield, Lanham (Maryland).

\qparr
Roehner (B.M.), Syme (T.) 2002: Pattern and repertoire in history. 
Harvard University Press, Cambridge (Mass.).\qL
[An updated version is available at:
http://www.lpthe.jussieu.fr/$ \sim $roehner/prh.pdf]

\qparr
Roehner (B.) 2007: Coh\'esion sociale. Odile Jacob, Paris.

\qparr
Rud\'e (G.) 1955: The Gordon riots. A study of the rioters and 
their victims.
Transactions of the Royal Historical Society, 11 June 1955, p.93-114.

\qparr
Rud\'e (G.), Harvey (J.K.) 1964, 2000: Revolutionary Europe,
1783-1815. John Wiley and Sons.


\count101=0  \ifnum\count101=1


\centerline{\bf \color{blue} Table of contents} 
\vskip 2mm

{\bf \color{blue} Single events versus clusters of similar events} \qL
A mysterious event \qL
From anthropomorphic to scientific understanding 
\qpar

{\bf \color{blue} Testing quantitative and qualitative models} \qL
Are there in the social sciences laws in the sense of physics?\qL
How to test qualitative models in physics?\qL
How to test qualitative models in history? 
\qpar

{\bf \color{blue} Requirements for a scientific analysis of historical 
         events} \qL
Complicated episodes should be decomposed into their 
        components \qL
One should study clusters of similar events \qL
One should focus on cases for which the effect is strongest 
\qpar

{\bf \color{blue} Further recommendations} \qL
Use broad sources and databases \qL
Prefer primary sources \qL
Avoid anthropocentric reasoning \qL
Ideologies in Durkheim-like view
\qpar

{\bf \color{blue} From physics to cliophysics} \qL
Distinctive features of the development of physics \qL
Guidelines for the development of cliophysics 
\qpar

{\bf \color{blue} What will be the future of cliophysics?} \qL
Reasons for pessimism \qL
Astrology $ \rightarrow $ astrophysics $ \Rightarrow $
         history $ \rightarrow $ cliophysics \qL
Some examples of cliophysical investigations \qL
The temptation to omit external factors \qL
Promises of cliophysics 
\qpar

{\bf \color{blue} What is to be expected from cliophysics?} \qL
Limitations of physics \qL
Limitations of cliophysics 
\qpar

{\bf \color{blue} Appendix A.
Is the quantity theory of money predictive?}
\qpar

{\bf \color{blue} Appendix B. Communist resistance to German occupation,
May 1940-June 1941}
\qpar

{\bf \color{blue} Appendix C. Practising cliophysics: 
two simple problems that can serve as initiations}
\qpar

{\bf \color{blue} References}

\vfill\eject

\large

\count101=0  \ifnum\count101=1

\qI{Testimonies}

Several co-authors of the present paper have had personal contacts
with Prof. Dietrich Stauffer. They briefly recall them in the following
testimonies.
\vskip 4mm

\qdec{\it \normalsize
I knew Dietrich very well when I worked at the RWTH 
(Rhine Westphalia Technical University) in Aachen 
from 1993 to 2000. We had coauthored a paper. My research about the critical
phenomena in finite systems above the upper critical dimension
was inspired by his questions. 
He supported my application 
for a professorship at the Institute of Theoretical Physics of 
the Chinese Academy of Science.}
\hbox to 15cm {\hfill \it Xiaosong Chen}
\qpar

\qdec{\it \normalsize
 I met Dietrich
during a visit to Wroclaw when we were both invited to dinner at the house of
Stanis{\l}aw Cebrat. During dinner it became clear Dietrich was thinking about
retirement and I enquired what he planned to do. I was anticipating like many
theoretical physicists, he would say continue working on aspects of sociophysics.
But he replied he planned to study the Second World War and learn where
Germany went wrong. Later on, he
spent a few days in Dublin and during his
visit I was impressed by his request to spend time with the students and postdocs
who plied him with questions all of which he patiently dealt with. 
}
\hbox to 15cm {\hfill \it Peter Richmond}
\qpar

\qdec{\it \normalsize It is in 1995 through Didier Sornette that I came
to know Dietrich. Although computer simulation was not really
my ``cup of tea'' we remained in fairly close contact over the
years. Given my interest in history I was particularly interested
when he told me that after retirement he wished to study
history (see more about that in an upcoming section).}
\hbox to 15cm {\hfill \it Bertrand Roehner}

\vfill\eject

\fi

 
Clio is the Greek goddess of history and cliophysics
is a neologism built
on the model of ``astrophysics'' or ``geophysics''.
It designates the study of historical events 
with the methods of  physics and more precisely of 
experimental physics.

\qI{Single events versus clusters of similar events}

If we say that single events can only be understood in an
anthropomorphic way whereas clusters of similar events can
be understood in a scientific way, this statement will
probably appear fairly unclear to many readers. 
That is why we start by illustrating that important
distinction through a real historical case.\qL
At the same time this example will allow us to mention
some of the  vivid discussions we have had with
Prof. Dietrich Stauffer, to whom
this special issue is dedicated.

\qA{A mysterious event}

Dietrich Stauffer, one of the founding fathers of
econophysics and sociophysics, manifested a great interest
for history, particularly after his retirement. 
For instance, he was much surprised to learn
that on the night from Saturday 26 to Sunday 27 of August
1944 
the German ``Luftwaffe'' conducted a bombing raid on Paris,
followed in the early morning by an attack of fighter planes
which straffed some of the streets of Paris%
\qfoot{See ``Chigago Daily Tribune'': 28 Aug 1944, p.3;
``Los Angeles Times'': same date, p.2; 
``New York Times'': same date, p.1.}%
.
Prof. Stauffer's astonishment was due to the fact that
this was a very special night. It was two days
after Paris had been liberated from German occupation by
a French and an American armored division and this very night
followed a great celebration of the liberation led by 
General de Gaulle on the Champs Elys\'ees and in which 
several hundred thousand Parisians took part.
Prof. Stauffer was so convinced that the Allies had total
air supremacy that, at first, he could hardly believe
that such a bombing really happened. Then, there was
a second thought which crossed his mind. 
Whereas these night raids in fact
attracted little attention (although some hundred persons 
were killed and several fires started) 
why were they not conducted a few hours 
earlier when the Champs Elys\'ees Avenue was crowded 
with so many people? Clearly, that would have
been a disaster and a great setback for the Allies%
\qfoot{There have been many accounts and also movies 
(e.g. Lapierre and Collins 1964)
about the liberation of Paris but the bombing seems
to have been ignored. 
Conducted by the ``IX FliegerKorps'',
the raid involved 111 aircraft, mostly Junkers 188 bombers,
in successive waves of 20 from south to north. The 
bombs killed 189 people, destroyed 430 buildings but 
avoided all historic places like the ``Louvre'' or the
``Arc de Triomphe'' which would have been easy targets
(Marchand 1985). }%
.

\qA{Anthropomorphic versus scientific understanding}

Why did we mention this event? \qL
The answer is simple. Although mysterious in several 
respects, this episode {\it cannot} be analyzed in a 
{\it scientific way} simply because it is 
a {\it \color{blue} single} event.
To get a better understanding, all we can do is to
collect more historical evidence. For instance, one
may ask where the bombers came from, whether
they were identified by Allied air defense radars
and, if so, why there was no interception.
Another interesting question is whether the obvious 
lack of air superiority over Paris was limited to 
night attacks 
or also extended to daylight attacks.
\qpar

While such answers may in a sense provide 
a better understanding, they will certainly not
lead to a {\it law} permitting predictions.
The kind of understanding they allow can be called
an {\it \color{blue} anthropomorphic understanding} in the sense
that it is based on a cascade of human reactions.
We use the term ``cascade'' because this kind of explanations
come about in successive steps.
For instance, if we learn that the bombers
were identified but no interception was ordered, in the
next step one may ask what was the reason of that decision.
Clearly, each answer will raise new questions in an endless
chain.
\qpar

During World War II,
apart from Paris, several other capitals were liberated 
from German occupation by Allied forces: Rome (4 June 1944),
Brussels (4 September 1944), Athens (15 October 1944),
Amsterdam (8 May 1945), Copenhagen (8 May 1945).
If at least some of these liberations had also 
been followed by a night raid, there would be a pattern
of recurrent events
which in turn could be the starting point 
of a scientific explanation. As far as we know, it is not
the case which means that the Paris bombing will remain
a single and isolated event.

\qI{Testing quantitative and qualitative models} 

As we said above, the hallmark of scientific analysis 
consists in the successful testing of laws. These laws,
we will argue here, can be quantitative as is usually the
case in physics but can also be qualitative as is often the case
in cliophysics.

\qA{Are there in the social sciences laws in the sense 
of physics?}
 
Quantitative laws will state that, under a number 
of specific conditions, one shall observe an effect 
whose magnitude is predicted by the law.\qL
The curvature by the Sun of light rays originating from 
distant stars, as predicted
by Einstein's General Relativity is a well-known illustration.
\qpar

Finding such laws is an ambitious objective which can only
be achieved for phenomena which are simple enough in the
sense of depending upon a small number of parameters.
As an illustration, 
in Appendix A we discuss the case of a long standing
economic law. Is 
the quantity theory of money (already stated by 
David Ricardo in 1810) a law in the 
meaning defined above? It will be seen that it
has indeed some predictive power but only in
situations of massive increases of the money supply,
for instance those which occur in time of war. The moderate
variations of the money supply seen in peacetime
appear too weak to trigger a change in price level
that clearly stands out from the background noise
(consisting in inherent price fluctuations).
\qpar

However, it would be a mistake to focus exclusively
on quantitative models. Social and historical mechanisms 
can often be described by qualitative models which 
can be tested quite as effectively as quantitative models.

\qA{How to test qualitative models in physics?}

In this and the next subsections we explain how 
experimental comparisons can be used to test
qualitative models first in physics and then in the
social sciences.
\qpar

Suppose you are living in
the time of Galileo (1564-1642) and
Descartes (1596-1650) and you ask yourself
what are the forces involved in the phenomenon of free fall.
A first idea is to try Galileo's method of dropping 
balls of different sizes and densities from the tower of Pisa.
However, with the poor means of observation available in that
time, this method will not work. \qL
Another approach is to try clever comparisons,
for instance by observing and analyzing the
fall of (a) steal balls (b) soap bubbles (c) feathers. \qL
As one knows, there are 3 forces at work. (i) the weight $ mg $
(ii) the buoyant force (iii) the friction of air. 
In (a) one sees almost only (i). In (b) the buoyant force (ii)
is magnified and by trying bubbles of
different diameters, one can identify its characteristics. 
In (c) it is the friction of air which is amplified and
by trying feathers of different sizes, one can identify
its characteristics. In his publications Galileo does not
report that he used such a comparative method 
although it would  have 
given him a better understanding of the phenomenon of free fall.
\qpar

Note that the description given above is a simplification for
there is also the Coriolis force due
to the rotation of the Earth and the ``added mass force'' 
discovered by Friedrich Bessel in 1828 and which results from 
the fact that the ball transfers a fraction of its
acceleration to air molecules.\qL
Incidentally, this
shows that even a simple phenomenon like free fall
must be decomposed into its diverse components to
become intelligible. We come back to this point later on.

\qA{How to test qualitative models in history?}

The idea is quite simple. It often happens that a 
historical question which 
cannot be settled clearly at the level of one country 
can be cleared up very easily when the field of observation 
is enlarged to similar cases in
neighboring countries. As an illustration
we focus on an episode  which occurred during 
the Second World War. 
\qpar

In France a great debate has been raging for decades
(and still resurfaces now and then) which concerns the attitude
toward Germany of
the French Communist Party (PCF) in 1940-1941. 
The PCF was a member of the
Comintern (also called ``Third International'') which is
a federation of Communist parties from across the world.
A common and fairly natural
belief is that, just as NATO is basically controlled by
its most important member, the United States, so the Comintern
was controlled by its most important member, the USSR.
The question is whether this was a rigid or a flexible 
control. 
\qpar

In the first case one  would expect all Communist 
parties in European countries occupied by Germany to start 
resistance actions only after the German attack on the USSR
on 22 June 1941  (remember that the non-agression pact between
Germany and the USSR was signed on 23 August 1939). \qL
On the contrary, if one can identify resistance actions before
June 22 it will support the thesis of a flexible control.
In short, the challenge here is to decide which one of the
two models, rigid or flexible,
is correct. Although the models are qualitative
(or, if one prefers, of the dichotomous $ 0-1 $ type) 
they are perfectly well defined. 
\qpar

The historical evidence is summarized in Appendix B. 
It turns out that whereas in France early resistance 
actions (i.e. before June 22) involved few people and
can therefore be denied by some, in the Netherlands
early resistance by the Communist Party of the Netherlands
involved thousands of striking workers
and therefore can hardly be ignored.
\qpar

More broadly, in the past 60 years there have been
recurrent debates in France about many aspects of the 
German occupation but, to our knowledge, there has
not been a single attempt to establish comparisons with
other occupied countries. A comparison with Denmark would 
be particularly enlightening because, as in France, an official
government remained in the occupied country; in most other
countries a government in exile was established in London.
\qpar

The previous example shows that comparative
analysis may solve fairly easily problems which otherwise
cannot be settled conclusively.
Such a neglect of the power of comparative history 
is all the more
surprising in a country like France which has had 
several renowned pionneers in comparative history, 
e.g. Marc Bloch, Pierre Chaunu and Fernand Braudel.

\qI{Requirements for a scientific analysis of historical events}

The requirements listed below are not different from
those commonly accepted in any experimental exploration
done in physics. In fact, these conditions
are not special to
physics but common to all sciences. The reason 
is that they derive from an crucial imperative of 
{\it simplification}. 
The real world is inherently complicated%
\qfoot{Note that we do not say ``complex'' for we
keep this word for systems whose elements have many 
long range interactions. It is true that, due to symmetry
properties, a system in which all elements interact
identically with all others is fairly easy to model
mathematically but, leaving aside this specific case,
in a general way systems with more interactions are capable
of rich and well structured responses. 
The immune system, the genome or 
the brain are complex systems
whereas the stock market is a complicated system in the sense
that there are many kinds of investors (as well as 
a great diversity of companies) but each investor
interacts with only a limited number of other investors;
moreover, the high level of noise precludes the
emergence of stable and well structured responses.}%
.
If we wish to {\it understand} it we cannot tackle it head on.
To split a rock or a diamond one needs
to identify its fault lines. The human mind cannot develop
an understanding from models which contain too many parameters.
Unless one renounces to any intuitive understanding (which is
is what is done in econometrics), simplification
is essential.

\qA{Complicated episodes should be decomposed into 
simpler components}

In the title 
of the paper we did not write ``a scientific analysis
of history''. Instead, we used the word ``event''.
The reason is simple. Our objective is not to understand
history globally. For instance,
it is impossible to analyze scientifically the French Revolution
of 1789 for it has many aspects:
the meeting of the Estates-General, the confiscation of
the property of the Catholic Church, the storming of 
the Bastille and of the king's palace (namely
the ``Tuileries'') in August 1792, and so on.
Clearly these are very different facets and they 
must be studied separately. 
\qpar

This decomposition prerequisite is the same as 
in physics. Usually one tries to avoid
experiments in which different effects are mixed up.\qL
This requirement is a direct consequence of the simplicity
imperative.

\qA{One should study clusters of similar events}

In the previous sections we have already discussed
a  prerequisite
that all scientists know very well but that one
is often tempted to forget, namely that there can be
no scientific analysis of single events. That is 
why in the title of the paper
we did not write ``of historical events''
but ``of recurrent historical events''. Single events can 
only be analyzed in what we called an anthropomorphic way.
\qpar

The word ``similar'' in the title of this subsection
makes reference to a specific mechanism (in physics
we would say a specific ``effect''). Sometimes
this mechanism can be defined easily.
For instance, the confiscation of Church property
is a well defined effect. 
In the time of the Reformation
it was common for sovereigns who became Protestant to
confiscate the estates of the Catholic Church. 
\qpar

However, there are also situations in which the 
mechanism which is at work is not immediately apparent
and will emerge only gradually through a comparison 
of various historical episodes.

\qA{One should focus on cases for which the effect is strongest} 

In principle the effect of buoyancy on free fall can 
be observed when a lead ball is replaced by as steel
ball, but in reality the difference is so small
that the effect will be lost in the background noise.
\qpar

In Appendix A we show that changes in the 
money supply will visibly affect the price level only 
if they are massive enough. 
This rule should be kept in mind when one selects
historical cases. It is a waste of time to study 
second order effects or cases in which the main
variable changes by less than 5\%. \qL
Here is an illustration. 
\qpar

In Richmond et al. (2018) it was shown that
a sudden death spike in a population
is usually followed 9 months later by a trough in the
birth rate.
For instance a terrible famine in Finland in the fall of 
1867 led to a sharp fall of the birth rate 9 months later
due to the fact that difficult living conditions
reduced the conception rate.
Why did this effect not attract earlier
attention? It is certainly because it can only be observed following
massive death spikes (e.g. the influenza epidemic
of Oct-Nov 1918 is another example). 
For less massive death spikes, the birth
rate trough is masked by the natural fluctuations
of the birth rate.

\qI{Further recommendations}

Although less crucial than the previous requirements, the
following recommendations will facilitate cliophysical analysis.

\qA{Use broad sources and databases}

Presently, just by browsing through historical papers
one comes immediately to the conclusion that historians
study mostly the history of their own country. It is fairly
clear that in such conditions the cliophysics approach 
that we develop
in this paper is doomed. The reason is obvious. This
approach can only work if one has access to a wide collection
of cases from which one can then select those which 
seem particularly appropriate for well-focused 
scientific studies. To limit oneself to the events of
one's own country is similar to astrophysicists
who want to understand how stars work but would limit
themselves to starts similar to our Sun, thus
ignoring red giants, blue stars and all
intermediate cases of the Hertzsprung-Russell diagram.
\qpar

Nowadays, thanks to the Internet, to the development of
digitized newspaper archives (e.g. those produced by ProQuest),
to the fact that many old books are available online,
the conditions are fulfilled which allow the broad
investigations advocated in the present paper. However, there
is still room for improvement; presently (in 2020) less
than ten major newspapers, mostly American, have digitized
all their archives.
\qpar

Apart from the Internet there has been another revolution.
The language barrier which for centuries has been 
a formidable obstacle has nearly disappeared. 
Over the past
5 or 6 years the quality of the translations
has improved dramatically to the point of
becoming quite acceptable%
\qfoot{An easy way test of a translation software
is to perform a ``circular translation'': 
$ A \rightarrow B \rightarrow C $. With a good software,
$ C $, although not identical to $ A $, has the same meaning;
as if the software was able to catch the meaning!}%
.

\qA{Prefer primary sources}

A last consideration can be added.
Historians make a distinction between
primary and secondary sources. Whereas primary documents
are supposed to be written by the very persons who took
part in the events, secondary sources are written by historians
based on what they read in primary sources.
What would be the analog in physics of relying on secondary
sources? 
\qpar

When you use primary sources it is like doing the 
experiment  yourself. When you use
secondary sources it means that you tell another person
(namely the historian) what experiment you wish him
to perform. Then,
when you get the results you can never be sure
how the experiment was really conducted. 
In other words, this
adds a layer of uncertainty to the whole process. 
What you get is a mixture of the primary account
and the feelings and biases of the historian.
What the historian records (or leaves out)
depends very much 
upon his own interests and opinions.\qL
In short, whenever possible it is always better to work
with primary sources. Fortunately,
in the future more and more archive sources will become 
available online. Naturally, the persons who took part in
the events are rarely objective but their prejudice is
usually easier to identify than the bias of the 
``sanitized'' accounts of historians.

\qA{Avoid anthropocentric reasoning}

In his book ``The rules of sociological method'' the
French sociologist Emile Durkheim (1894) takes great care
to emphasize that social events should be studied
like ``things'' (``comme des choses'' in the French text),
in other words they should fight anthropocentric
temptations.
What does he mean with this advice? Because social
scientists are human beings just as the people 
whose behavior they study they may be tempted to
substitute their own thoughts to those of these people.
\qpar

Here is an illustration in the form of a dialog. \qL
``Why do people commit suicide? Well, because they are
sad and depressed. \qL
In which season are people most likely 
to be depressed? Well, in winter time when the days 
are short  and nature is at a standstill''. \qL
This sounds plausible enough. \qL
Yet, in the northern hemisphere suicide rates are minimum
in December and maximum in May. Surprising, isn't it?
\qpar

As this is an important point let us give another illustration.
In June-July 2020 there were daily demonstrations in Portland, Oregon.
Why?
A possible answer may involve explanations about the attitude of
President Trump, the ``Black lives matter'' movement and so on.
Clearly, these are anthropomorphic explanations. To get
a Durkheim-like view one should watch these demonstrations
with the eyes of a ``Martian''. He does not understand what the
demonstrations are about but he can see that over past decades
there were recurrent demonstrations in Portland
(Wilson 2019). Having observed that special feature,
our Martian naturally expects that any new spark will trigger
more demonstrations. In other words, he will be led to
a correct prediction without knowing any of the idiosyncrasies
of the situation.

\qA{Ideologies in Durkheim-like view}

Ideologies (e.g. Christianism, Islam, Communism, anti-Communism,
neoliberalism and so on) have played
and continue to play an important role in history. How can we represent
them in a Martian-like perspective? Very simple. We need only to be
able
to estimate the strength of the inter-individual attraction created
by these ideologies and this can be done by setting up some appropriate
{\it sensor}, e.g. the percentage of people who attend religious service.
In other words, the strength of the ideologies is sufficient,
we do not need to bother about their contents,
that is to say whether the persons
are Baptists, Catholics, Mormons or Quakers.

It is important to realize that the challenges and difficulties
faced in cliophysical studies were also present
throughout the development of physics.

\qA{Distinctive features of the development of physics}

Physics, so far the most successful science, 
was graced with three favorable circumstances.
\qee{1} Over several centuries physicists did not have
the experimental means to explore 
the microscopic structure of matter.
Thus, willy nilly, they had no other choice
than to focus on the laws of {\it \color{blue} macroscopic physics}.
Would they ever have discovered the simple law which rules
the refraction of light if they had started from
the atomic level%
\qfoot{Finding a simple microscopic definition of the refractive
index would already be a serious difficulty.}%
?
\qee{2} In scientific observations the main obstacle 
is what is called the {\it \color{blue} background noise}.
High background noise reduces the signal-to-noise ratio
and makes the detection of patterns and laws more difficult.
Significantly, the discovery by Kepler around 1600 of the 
three laws which rule
planetary orbits was made on a system (namely
the trajectory of Mars around the Sun) subject to little
background noise. Then, in the following centuries,
physicists moved gradually to the study of systems
with ever larger fluctuations. Hydrodynamics with its
inherent turbulence was a difficult challenge and after
that came microscopic quantum phenomena 
for which the very notion
of well defined deterministic trajectories disappears.
\qee{3} Finally, in each of its new fields, 
physics started from
the {\it \color{blue} simplest possible} system: the two-body case
before the three-body problem, the ideal gas before
real gases, the hydrogen atom before lead or uranium atoms. 
\qpar

Why did we recall these milestones in the development
of physics? The reason is simple.
We are convinced that they provide 
valuable guides for the development of cliophysics.
This is explained in the following subsection.

\qA{Guidelines for the development of cliophysics}

The previous observations give the following guidelines
for a sound development of cliophysics.
\qun{Macrocliophysics} The first of the points 
mentioned above teaches us to start by studying
the laws of macrocliophysics.\qL
At first sight this may seem somewhat counter-intuitive. 
As the ``atoms'' of cliophysics
are human individuals we are in a situation 
completely opposite to that of physics.
Not only do we know these ``atoms'' but we are part of them.
Naturally enough, one is tempted to use our
knowledge of humans 
to explain social
phenomena through the characteristics of individuals%
\qfoot{Paul Lacombe (1894) has tried to develop
a science of history based on the psychological
features of individuals but its attempt led
nowhere, which is quite understandable because
ultimately psychology relies on an understanding
of the brain, a system of overwhelming complexity.}%
. 
That would be a mistake, however.
Statistical physics has been developed only fairly late
in the history of physics and it is still a challenging field.
Can we predict from the properties of water molecules
that its boiling point is 100 degree Celsius?
For cliophysics the difficulty is even much
greater because of the multiple degrees of freedom 
of individuals.
\qun{Noise reduction} The second of the points made 
above can explain why
cliophysics has not yet been developed. 
Socio-historical phenomena
are very diverse which means a high background
noise unless it can be reduced by a smart selection of
the phenomenon under observation.
Astrophysics teaches us how to do that.
Clearly,
the trajectory of Jupiter follows the three Kepler laws
with much higher accuracy than the trajectories of
asteroids for the latter are subject to the 
gravitational attraction of the planets which come
across their orbits. In the same way the international
relations of a small country will be subject to more
interferences and perturbations than those of a major power.
\qun{Simple cases}
The third of the points made above suggests that we should
start by studying the simplest possible systems;
``simple'' here means involving a small number of
factors. This seems obvious enough but often it
comes in conflict with our personal interests.  
As citizens and members of certain communities
(whether religious, linguistic or ethnic) we have specific
interests. The main imperative of cliophysics is
that in our investigations we must avoid such
anthropocentric tendenties%
\qfoot{We use the word ``anthropomorphic'' to
mark the difference between 
humans and non-human (including non-living)
entities. The word ``anthropocentric'' refers more
narrowly to the personal interests of individuals,
for instance people may be more interested in 
their own country than in others.}%
.
To convince ourselves that this is a crucial
point one needs only to recall that in the historical
development of science the temptation to put 
humans in central position has been a major obstacle.
Geocentrism or the rejection of the evolution of species
are classical examples. This anthropomorphic attitude
is still present in our time.

\qI{What will be the future of cliophysics?}

\qA{Reasons for pessimism}

In 2002 one of the co-authors (BMR) published a book
which already contained several of the ideas explained in the
present paper. Then, two things happened, each of which was 
rather disappointing.
\qbu In the wake of the publication there were a number
of reviews. One of the reviewers who understood very well
the purpose of the book came up with the following 
argument. 
\qdec{``You say that when you have several similar recurrent
events you can predict that the outcome of episode number $ n+1 $
will be basically the same as the outcomes of the previous
$ n $ cases. \qL
Well, let us apply this rule to your project.
In the past, there have been already several attempts
to make history more scientific. As far as we know,
they all failed. Thus, can we not conclude that your 
project will also be unsuccessful?''}
\qpar

Naturally, this conclusion holds only if all other 
conditions remain more or less unchanged
\qfoot{Some two thousands years before Kepler,
astronomy had already emerged from astrology thanks
to a number of Greek scientists who designed
and carried out wonderful experiments. Then, after
Claudius Ptolemy (100-170) further progress stopped.
Ptolemy's tables were only used by astrologists and
geocentrism was held as the only possible truth
compatible with the Bible. In other words,
first in Greece and then in western Europe
favorable ``conditions'' (which ones we do not
know exactly) occurred which
allowed the emergence of astronomy.}
. 

This was a clever argument. However, it holds only
as long as there is no change of paradigm. The fact that
successive geocentric models were unsuccessful does not mean 
that a heliocentric model will also fail. We believe that
the transition to cliophysics represents a change of
paradigm. 
\qpar

Nevertheless,
even if the previous argument does not apply, one must
recognize that what happened in
the 18 years since the initial publication confirmed that
sad prediction in the sense that this approach
attracted no attention and was not used by other researchers.
\qpar

In a sense this is hardly
surprising. One century ago, in the time of social scientists 
like Emile Durkheim (see Durkheim 1894), 
Vilfredo Pareto (see Pareto 1919) or Marc Bloch (see Bloch 1924),
such a project may have been understood for in that
time the methodology of physics served as a model for
social scientists.  After the Second World War the situation
changed completely for instead of physics, American social
scientists started to take econometrics as their model
and they were soon followed by the rest of the world%
\qfoot{Apart from the number of parameters and variables, 
econometrics also differs from physics in how it uses
statistics. There are two distinct approaches in 
mathematical statistics.
One is based on confidence intervals and the other
on significance tests. Physics uses the first which has
the advantage of
a direct connection with the notion of measurement 
accuracy  whereas
econometrics uses the second (one, two or three stars).}%
.
Yet, whereas physics is a highly successful field,
econometrics turned out to be a failure (except perhaps
for short-term management). How can one get an 
understanding from a model which has 20 parameters?
\qL
As a result, earlier innovative and
promising attempts, like the one by Pareto mentioned
above, were completely forgotten.
\qpar

So, is there no hope? In the following subsection
we explain why there are nevertheless some reasons to 
be optimistic.


\count101=0  \ifnum\count101=1

\qA{Astrology $ \rightarrow $ astrophysics $ \Rightarrow $
history $ \rightarrow $ cliophysics}

Before we start our explanation it should be realized that
technically astrology is a very sophisticated field.
From China to the Middle East and to Western Europe it has
been in existence for over 2,000 years. Tycho Brahe and
Kepler, two founding fathers of modern astronomy, served
as astrologists of the Emperor of the Holly Roman 
Empire (i.e. present day Germany). One can understand 
that astrology was an attractive discipline for its purpose
was really to help people. What can be more useful 
for an emperor than to know on which day he should wage 
a battle in order to have the best chance of victory? \qL
In contrast, the precise observations done by Tycho Brahe 
appeared  fairly useless because in order to locate
planets in their constellation (which was the main requirement
for astrologists) low accuracy was sufficient.
The three Kepler's law were published around 1609 and one
would have to wait some 80 years until 1687 for Newton
to show that they were compatible with his law of gravitation.

\fi


\qA{Some examples of previous cliophysical investigations}

It is likely that in order to become convinced by the approach
of cliophysics, our readers need to see that it works.
The best way would be to try it themselves by applying
this approach to a question in which they are interested.
However, before devoting some time to such an investigation
they may wish to see to what extent previous investigations
were successful.  This leads us to list some of the studies
performed during the past three decades. Instead of
listing a large number of studies just by name,
we prefer to explain in some detail just a few.
Other studies can easily be found on Internet
by using as key words the names of some of the co-authors.
\qbu The ``Central Intelligence Agency'' (CIA) was created in 1947
to fight Communism worldwide. Similarly, 
the Catholic order of the ``{\bf \color{blue} Society of Jesus}'' (SJ) 
was created in 1540 to fight the Reformation in all
European countries. The objectives were similar,
namely to get close access to political leaders in order to
influence their decisions and to weight on public opinion
through a control over teaching institutions%
\qfoot{This activity was not limited to Europe.
Founded in 1789,
Georgetown University remains a prominent Jesuit school
although it is also open to non-Catholic
students including 400 Muslim students (as of 2007).
The school's alumni include more US diplomats 
than any other university. For instance,
CIA director George Tenet was a
Distinguished Georgetown Professor in diplomacy.
Georgetown has also a Liaison Office at Fudan University in
Shanghai.
For more details see the corresponding Wikipedia articles.}%
.
Yet,
there was also a crucial difference in the sense that, at least
in Catholic countries, the Jesuits were working openly.
In a number of cases the Jesuits went too far
with the result that there were several expulsions
not of individual Jesuits but of the whole Society
from Catholic countries. These expulsions (almost 20
spread over 4 centuries) were compared
side by side in order to identify common features
(Roehner 1997a).

\qbu {\bf \color{blue} Separatism} is a phenomenon 
that has existed in
all times. The fact that there have been so many 
separatist uprisings makes it an excellent topic
for comparative analysis. As a matter of illustration
let us briefly describe two cases. \qL
In 60 Queen Boudica led a 
rebellion in Britain against the military occupation
by the Roman army. According to the two Roman historians
Tacitus and  Dio Cassius, this uprising which occurred 
some 20 years after the conquest was largely a result of 
the arrogant behavior of Roman veterans who 
had received land grants. At first successful, 
the rebellion was quashed within a few months by
the Roman legions. \qL
A more successful secession was the rebellion in 1581 of
the 7 United Provinces of the Netherlands against
Spanish occupation. The rebellion occurred
some 25 years after the region had been taken over 
by Spain and it is likely that here too the arrogance of 
the Spanish soldiers was a factor. \qL
A broad study of many separatist movements brought
to light the main factors which are at work in a
long range perspective 
(Roehner 1997b, Roehner and Rahilly 2002). In physical
terms a separatist movement is like a drop of oil in
water. The oil drops do not mix because the oil-water interactions 
are smaller than the oil-oil and water-water interactions.
This may seem a rather schematic view but in fact it gives
clear political guidelines. For instance, increasing the fare
and reducing the frequency of the ferry between Corsica
and continental France can only boost separatism.
\qpar

The previous explanation in terms of social interactions
can be seen as a microsocial model. 
A macrosocial factor is the strength of the central government.
History shows that a weak central government favors separatism.
An extreme case was the Holly Roman Empire  which comprised
some 300 independent entities: kingdoms, principalities,
free cities, and so on.
The European Union faces
the same risk for all 28 governments have willingly transferred
some of their powers to Brussels, but unfortunately
the European
Commission remains a very weak and unpopular form of government.
This opens the door for regional fragmentation of the kind
one can already see in Catalonia.
\qpar

If a weak state favors separatism, the converse is
also true in the sense that successful
separatist movements make states more fragile.
That is why
in the past 4 or 5 decades separatism 
has become a major political
weapon. One recalls that the secession of Lithuania
from the Soviet Union in 1989 gave a signal which led within
a few months to the secession of the other Soviet Republic%
\qfoot{It is often said that it was a disintegration
but this is not correct for each Republic had the
right to proclaim its
independence, as attested by the fact that Ukraine
and Belarus were already United Nations members. On the
contrary, a province like Chechnya was an ``autonomous republic''
within Russia which did not have the same status as the 
republics composing the USSR; it is true that
usage of the term ``republic'' in both cases does hardly
highlight this important distinction.}%
.
In the studies mentioned above we left aside all cases
in which there was obvious foreign interference for we wanted
to to restrict ourselves to endogenous factors. Nowadays
(in 2020) it would be almost impossible to find
separatist struggles not dominated by exogenous factors.\qL
As a case in point, the US Congress has passed three successive
acts which define and codify US interference in Tibet: \qL
(i) The ``Tibetan Policy Act'' of September 2002. \qL
(ii) The ``Reciprocal Access to Tibet Act'' of December 2018.\qL
(iii) The ``Tibetan Policy and Support Act'' of January
2020. This last act is about the succession of the Dalai Lama.
\qpar

To study the case of Tibet without taking into account
the exogenous factors would be similar to  measuring
the period of a pendulum outdoor in strong wind without
taking into account the effect of the wind. On the other
hand it is very difficult to integrate the exogenous factors
into a model because we ignore how precisely the
measures defined in the acts were (and will be) implemented.
\qbu {\bf \color{blue} Fragmentation} is an effect that is
strongly related to separatism. The fragmentation of Yugoslavia
in recent decades shows that there is no ``natural'' limit to
such a process in the sense that it may extend to ever
smaller entities. Naturally, the successive
fragmentations (and re-unifications) of China has attracted 
considerable attention; see for instance the study of Baaquie and
Wang (2018).
\qbu {\bf \color{blue} US responses to challenges in the Pacific.}
Ever since the United States took possession 
of the West Coast (the conquest of California was in 
1847 during the US-Mexican War) it has had an
hegemonic position in the Pacific Ocean. As a matter of fact,
from 1850 to around 2010 it was by far the most important
power of the Pacific rim to the point that President
Eisenhover called it an American lake. In 
a book by Zengru Di et al. (2017) and in 
a paper by Belal
Baaquie et al. (2019) US responses to successive 
encroachments upon its hegemony from 1880 to now were 
systematically studied. It turns out that during this
time interval the US has been unwilling to consider
a negotiated partnerships preferring to give a free
hand to its military%
\qfoot{Yet, one can remember that in April 1951
President Truman
reined in General McArthur's in his plans against China.
In this plan ``between 30 and 50 atomic bombs'' would
have been  dropped on China, see the interviews
of MacArthur taken in 1954 but which were 
published in the ``New York Times'' ten years later on 
9 April 1964 (p.16).}%
.

\qA{The temptation to omit external factors}

In the case of Tibet external interference is quite clear
but this is in fact exceptional. Most often, when a country $ A $
tries to influence a country $ B $, neither $ A $ nor $ B $
wants to recognize that interference openly. 
For $ A $ it would suggest 
arrogance whereas for $ B $ it would imply weakness.
That is why a
common drawback of most historical accounts is the omission
of foreign
interference. This is a variant of the anthropocentric
temptation discussed above in the sense that historians are
so focused on their own country that they are led to
forget external factors.
\qpar

To make this point one could rely on fairly general
arguments. For instance, it can be observed that 
countries (e.g. Japan or Australia) which rely for their
defense on the US nuclear umbrella can hardly have an
independent foreign policy. Yet, instead of such broad
reasons it may be more convincing to describe a specific case.
\qpar

In May 1958 the political situation in France opened
an opportunity for General de Gaulle to come back to
power (remember that after having led France through
World War II he had resigned in 1946). This process 
which lasted
some three weeks from May 13 to early June is described in
almost all accounts as a purely French story. 
Yet, General de Gaulle
himself was quite aware of the importance of external factors
to the point that he sent an envoy to the US embassy in
Paris who pledged in his name that France would remain 
a member of NATO.  Although almost never mentioned, this contact
is duly recorded in the official archives of US
foreign relations as published by the State Department
(the complete collection of volumes is available on Internet).
Incidentally, this pledge may explain why France left
NATO only during the second 7-year presidential term of
President de Gaulle.

\qA{Promises of cliophysics}

Nowadays governments no longer employ astrologists 
but they do 
not have any analytical tool to help them take right
strategic 
decisions. There is admitedly a common belief that the
past can help to understand the present but this is 
taken as a
loose statement without any real usefulness. 
Some leaders like General de Gaulle or 
Winston Churchill had a good knowledge of the past 
and a ``sense
of history'' which naturally led them to establish parallels
with former situations. 
The comparative analysis presented in the present paper
systematizes what such leaders were doing more or less 
intuitively. The guiding principle can be summarized
in two sentences. ``Identify former episodes which display 
the same mechanism. Then, what has happened many times is 
likely to happen again''. When an astrophysicist wants to
study neutron stars he does not rely on only one observation,
but collects all available cases. The same should be done
for historical events. Because they are trained accordingly
astrophysicists and physicists have a special role
to play in the developement of this project. For many decision
makers cliophysics should be of great help.

\qI{What can be expected from cliophysics?}

\qA{Limitations of physics}

Physics textbooks entertain their readers in the wrong belief that
the laws of physics allow us to explain {\it all} natural phenomena.
In reality, there are many phenomena for which the laws of physics are
unable to offer successful predictions. Here are two examples.
\qbu Can one predict the shape, size and color of clouds?
\qbu Many solids exist in different allometric states. Can
physical chemistry predict the number and characteristics of these
states?
\qpar

In short, although the general framework is fairly well understood,
there are many questions which just turn out to be too challenging.

\qA{Limitations of cliophysics}

Why was it important to make this point at the beginning of a
paper about cliophysics? 
When scientists take a closer look at cliophysics, they come
quickly to realize that cliophysics (at least 
as developed in this paper) can be used successfully only
for specific classes of phenomena. In all other cases, just as
in the physical phenomena listed above, its application
would be too cumbersome and uncertain. 
\qpar

While honesty required us to emphasize such limitations,
we wish also to give our readers a more positive perception.
Subsequently we describe problems to which 
cliophysics has already been applied with 
some success; moreover in Appendix C we propose two
questions that we think suitable as
initiations for those
of our readers who wish to try this approach by themselves.

\appendix

\qI{Appendix A.
Is the quantity theory of money predictive?}

Just to give a sense of how difficult it is to find real laws
in the social sciences
we give an example from economics. What is the predictive
power of the quantity theory of money?
\qpar

This theory
states that there is a relationship between the
price level $ p $ and the 
amount of money $ M $ in circulation in an economy 
in a given year. For an economy based on a single product, $ A $,
and a single form of money, e.g. cell phone money, the
relation will be as follows:
 $$ Mv = pQ  \qn{1} $$
where $ v $ is the velocity of money, i.e. basically the 
annual number of transactions; $ Q $ is the quantity of $ A $
produced in one year. Clearly this equation can be
easily generalized to several products and various forms of money 
(cash, credit card, cellphone). According to equation (1),
if $ v $ and $ Q $ are
stable, doubling  $ M $ should double $ p $.
\qpar

This equation works fairly well when a massive amount of
printed money is injected into an economy. Below we 
mention several cases.

\qA{Massive postwar introductions of banknotes}

A case in point was the 
postwar occupations of Japan and South Korea by US troops
who brought with them a huge amount of bank notes (of the
order of the total receipt of the budget, therefore
equivalent to
a budget deficit of the order of 100\%) printed 
in the United States.
This led to a considerable inflation between 1945 and 1950
whose effects are still
visible nowadays through the many zeros printed on Japanese and
South Korean banknotes. 

\qA{Paul Volcker's battle against inflation}

A second episode can be mentioned which was a practical
application of the theory by Paul Volcker,
the Chairman of the US Federal Reserve. In March 1980
the US inflation rate reached 14.8\%. Volcker raised
the interest rate (more precisely the federal funds rate)
from 11\% in 1979 to 20\% in June 1981. The inflation
rate fell to 3\% in 1983 but the high interest rate had
adverse side effects: (i) a sharp recession
(ii) an overvaluation of the dollar which hindered
US exports. After only two terms from 1979 to 1987
Volcker was fired by President Reagan. Volcker's 
successor, Alan Greenspan, remained chairman from
1987 to 2006.

\qA{``Quantitative easing'' after the crisis of 2008}

A third episode is the period of massive money
``printing''  (also called quantitative easing) 
by the central banks in the wake of
the crisis of 2008. \qL
Some economists were surprised that it did 
not generate any substantial inflation. In fact, 
it did create inflation but which remained limited to the
prices of stocks and real estate. That is not surprising
because the supply of money by the Treasury
was injected into the economy at the
level of big banks (e.g. in the US the banks belonging to
the Federal Reserve system). It is natural to think
that for these banks the most natural way to use this money
in a profitable way was to inject it into the
stock market. An increase in stock prices makes everybody 
happy as it creates an optimistic economic climate.

\qA{Massive money supply during the Covid19 crisis of 2020}

A fourth test will be the aftermath of the Covid19 crisis
which gave rise to even more massive quantitative 
easing than the financial crisis of 2008. Will it
lift up $ p $ in accordance with equation (1)? 
\qpar

In conclusion one can say that, with the exception of the
spectacular occupation episodes, the quantity theory of money
showed limited predictive power.

\qI{Appendix B. Communist resistance to German occupation}

Before giving the evidence for France we will describe
the evidence for the {\bf Netherlands}.
In the Wikipedia article about the Communist Party
of the Netherlands one learns of the two following actions.
\qbu As early as 15 May 1940, the day after the Dutch capitulation, 
the Communist Party of the Netherlands (CPN) 
held a meeting to organize resistance against the German occupiers. 
It published a resistance newspaper called De Waarheid (The Truth).
\qbu Then, in February 1941, there was a general strike
against the arrest of several hundred Jews 
(United States Holocaust Memorial Museum, 
article ``Amsterdam''.
On 26 February, 300,000 people joined the strike. 
Broken by the Germans the strike lasted only three
days.\qL
However in later deportations in 1942 the Dutch municipal
police cooperated in the deportations (just as it did in France
and probably for the same reason).
For the whole
duration of the war, 75\% of the 130,000 Dutch Jews were deported. 
\qpar

For {\bf France} one piece of evidence is an article in the
New York Times of 21 October 1940, p.3 entitled:
``Three French reds arrested. Cache of arms and Communist
literature found in Lyon''. Another piece of evidence 
can be found in
a book by Alain Gu\'erin (1972) which cites sabotage actions 
by Communists in early 1941. \qL
Then, on 15 May 1941, the PCF set up
the ``Front National de la lutte pour la lib\'eration de la
France'' (National Front for the liberation of France).
Clearly, the actions in France were not very  
visible which is
probably why the debate could not be settled conclusively.
It is thanks to the Dutch case that a clear conclusion 
can be reached.

\qI{Appendix C. Practising cliophysics: two simple problems}

We have explained how,
following the guidelines of physics and avoiding
the trap of anthropomorphism,
cliophysics may be duly developed.
Then, we have listed a number of studies in which
this philosophy was put to work. However,
as nothing can replace personal experience,
we suggest here two test-studies  that are new and
can be tried by our readers if they wish.

\qA{The fate of head of states after revolutions}

After a revolution or a war how should the
new rulers handle the former head of 
state, referred here as $ H $ and assumed to be a male. 
 \qL
There are several options: (i) to expel him (ii) to try him
and then keep him in jail until he dies.
(iii) to sentence him to capital
punishment (iv) to execute him summarily that is to
say without any trial.
\qpar

In the course of centuries all these options have been tried.
Moreover, if one extends the investigation to the whole 
world, one can collect a substantial number of cases.
Cliophysics has the following message: ``Before taking any
decision, consider a number of past cases whose circumstances
were similar to those of $ H $.  \qL
Although very simple, such an advice has, to our knowledge,
never been followed; actually, has it ever been given?
\qpar

From the debates which preceded such momentous decisions, we know 
some of the arguments, e.g.
(i) ``We want a radical break with the past.''
(ii) ``We should take care not to create a martyr.''
(iii) ``A regime which starts violently will also end in violence.''
\qpar

Needless to say, all such statements are usually made without
any real historical basis. No attempt is made
to use the past to reach a smarter decision, not even when the 
cases are fairly similar. One can be fairly sure that the case
of Charles I sentenced to death by the English Parliament was
not brought up during the trial of Louis XVI by the French 
Parliament one century and a half later%
\qfoot{Not only were the charges similar but so was also
some twenty years later
the subsequent return to power of Charles II in Britain and Louis XVIII
in France.}%
.

\qA{The ``production'' of Loyalists}

The question of the American Loyalists during
the War of Independence has attracted much
attention from US historians. Those Americans who
sided with the British had their property confiscated
and eventually emigrated to other parts of the
British Colonial Empire.
\qpar

Needless to say, all great changes, whether political or
religious, produce Loyalists that is to say people who
wish to stick to the former conditions.
When Sweden became a Protestant country in the 16th
century, the people who remained Catholic were ``Loyalists''.
As they did not enjoy the same rights than Protestant citizens,
instead of remaining in the country some of them
fled to Catholic countries, e.g. the American
colony of Maryland. The same observation can be made
in Britain where Puritans, Quakers and members of
non-conformist denominations went to America
(Puritans to Massachusetts and Quakers to Pennsylvania).
\qL
Apart from religious revolutions one can also
consider  political upheavals, e.g. the Cromwell Revolution,
the French Revolutions of 1789 and 1792,
the Russian Revolution of 1917, the Chinese Revolution
of 1949. All of them produced Loyalists.

\qA{Religious loyalism in Japan}

Religious loyalism is probably less known than political
loyalism. It can be nicely illustrated by the case of 
Buddhism in Japan. \qL
Hardly known in the west is the fact that during the Meiji
Restoration there was an attempt to abolish Buddhism
altogether,
firstly by fueling distrust among the population and
secondly by detaching Shintoism from Buddhism. 
As a result, about one half of the Buddhist temples were
vandalized and destroyed with the complicity of the
authorities. Subsequently, Buddhists regained some
of their prestige by becoming an important political
pressure group. Additional information can be found
on the Internet through the keywords: ``Haibutsu kishaku'',
``Shimaji Mokurai'' and ``Jodo-shin sect''.
\qpar

Incidentally, the fact that in the west so little is known
about such important Japanese events  points out a major
difficulty of cliophysics. In the study of physical phenomena
our personal tastes play little role. On the contrary,
in cliophysical studies
our nationalist inclinations and cultural feelings may play
the role of a filter, at least unless we take great care and efforts
to set up a fairly ``universal view'', just as the Martian
mentioned previously would do.

\qA{How to start?}

An important question comes to mind immediately.
Should the investigation start with political or religious
cases? As already said, this study has never been done,
which means that we do not know. However,
one may think that the religious cases involve
fewer parameters for the obvious reason that a religion is
not expected to ensure the welfare of people.
Therefore one can ignore most connections with the
material world.
\qpar

Whereas at first sight the previous episodes look
exceedingly different, as often in science,
a preliminary classification may be useful.
Then, it may become possible to define
a few key-variables, e.g. the  proportion of people
killed, dispossessed or who emigrated.
\qpar

Once this is done, comes the interesting part
in the sense that it becomes possible to
study the main determinants of this
effect. Certainly, the duration of the previous
regime is of importance. A regime which had lasted only
a few months will hardly produce any Loyalists.
A second determinant may be
how strong were the links between the  previous regime
and the population.
Thus, little by little, a better understanding may develop
which in turn will allow us to ask sharper questions.

\vskip 10mm

{\bf References}

\qparr
Baaqui (B.E.), Wang (Q.-H.) 2018:
Chinese dynasties and modern China. Unification and fragmentation.
China and the World, Ancient and Modern Silk Road 1,1,1-43.

\qparr
Baaqui (B.E.), Richmond (P.), Roehner (B.M.), Wang (Q.-H.) 2019:
The future of US-China relations: a scientific investigation.
China and the World. Ancient and Modern Silk Road 2,1,1-53.

\qparr
Bloch (M.) 1924: Les rois thaumaturges. 
\'Etude sur le caract\`ere surnaturel attribu\'e \`a 
la puissance royale particuli\`erement en France et en 
Angleterre. Istra, Paris.\qL
Translated into English under the title: The royal touch.
Sacred monarchy and scrofula in England and France.
Routledge 1974.

\qparr
Di (Z.), Li (R.), Roehner (B.M.) 2017: 
USA-China: cooperation or confrontation. 
A case study in analytical history (in Chinese). \qL
[This book can be read at the following address:\qL
http://www.lpthe.jussieu.fr/$ ~ $roehner/prch.php]

\qparr
Durkheim (E.) 1894:  Les r\`egles de la m\'ethode sociologique.   
Flammarion,  Paris.\qL
[The book has been translated into 
English under the title:  \qL
``The rules of sociological method''.  \qL
Both the French and the English version are freely available 
on Internet.]

\qparr
Gu\'erin (A.) 1972, 2000:  Chronique de la R\'esistance.
Livre-club Diderot 1972, Omnibus 2000, Paris.

\qparr
Lacombe (P.) 1894:  De l'histoire consid\'er\'e comme science.   
Librairie Hachette, Paris.  \qL
[The following excerpt from the Preface suggests that the book
has the same objective as the present paper:
``Making history into a science  
would enable us to use with profit the overwhelming amount 
of recorded facts; once organized and linked to one another 
their weight will become more bearable''.
However,  
instead of following Durkheim's advice 
to study social phenomena like ``things'', the author
uses an anthropomorphic approach based on the psychology of 
the individual man. 
This leads nowhere because
the human mind is of much greater 
complexity than human societies.]

\qparr
Lapierre (D.), Collins (L.) 1964: Paris br\^ule-t-il? 25 ao\^ut 1944.
Histoire de la Lib\'eration de Paris. Robert Laffont, Paris.
An English translation was published in 1965 under the title:
``Is Paris burning? Adolf Hitler, August 25, 1944''. \qL
[The title refers to German orders to burn Paris but there
remains a mystery because
the unopposed German bombing in the night of
26-27 August 1944 did not target any historic building despite
killing 189 residents and leveling 431 houses.]

\qparr
Marchand (A.) 1985: Paris sous les bombes. En ao\^ut 1944 la capitale
\'était bombard\'ee par la Luftwaffe. [Bombing of Paris. In
August 1944 Paris was bombed by the Luftwaffe.]
Air Fan (French magazine), number 81, July 1985.

\qparr
Mathie (R.T.), and J. Clausen (J.) 2015: 
Veterinary homeopathy: meta-analysis of randomised 
placebo-controlled trials, Homeopathy 104,3-8

\qparr
Pareto (V.) 1919: Trait\'e de sociologie g\'en\'erale. Payot.
2 volumes.\qL
[The book has been translated into English under the 
title: The mind and society. 1935]

\qparr
Richmond (P.), Roehner (B.M.) 2018: Coupling between death 
spikes and birth troughs. Part 1: Evidence.
Physica A 506,97-111.

\qparr
Roehner (B.M.) 1997a: Jesuits and the state. A comparative 
study of their expulsions (1520-1990).
Religion 27,165-182.

\qparr
Roehner (B.M.) 1997b: Spatial determinants of separatism.
Swiss Journal of Sociology 23,1,25-59.

\qparr
Roehner (B.M.), Rahilly (L.) 2002: Separatism and integration.
A study in analytical history. 
Rowman and Littlefield, Lanham (Maryland).

\qparr
Roehner (B.M.), Syme (T.) 2002: Pattern and repertoire in history. 
Harvard University Press, Cambridge (Mass.).\qL
[An updated version is available at:
http://www.lpthe.jussieu.fr/$ \sim $roehner/prh.pdf]

\qparr
D. Stauffer (D.)  H.Eugene Stanley (H.E.) 1990: 
From Newton to Mandelbrot. A primer in
theoretical physics. Springer.

\qparr
Stauffer (D.), Oliveira (S.M. de), 
Oliveira (P.M.C. de), S\'a Martins (J.S.) 2006, 2012:
Biology, sociology, geology by computational physics. Elsevier.

\qparr
Wilson (J.) 2019: How Portland's liberal utopia became the center
of the right-wing war in the US.
The Guardian (newspaper), 16 August 2019.

\fi

\end{document}